\documentclass[twocolumn,aps,pra,floatfix,superscriptaddress]{revtex4-2}

\usepackage{graphicx}
\usepackage{dcolumn}
\usepackage{amsmath,amsthm,amssymb,physics,bm}
\usepackage{float}
\usepackage{cases}
\usepackage{multirow}
\usepackage{soul,xcolor}
\makeatletter
\let\newfloat\newfloat@ltx
\makeatother

\setstcolor{red}

\usepackage{algorithm,algorithmic}
\usepackage{hyperref}

\begin{document}

\title{Seeding Gaussian boson samplers with  single photons for enhanced state generation} 
\author{Valerio Crescimanna}
\affiliation{National Research Council of Canada, 100 Sussex Drive, Ottawa, Ontario K1N 5A2, Canada}
\affiliation{Department of Physics, University of Ottawa, 25 Templeton Street, Ottawa, Ontario, K1N 6N5 Canada}
\author{Aaron Z. Goldberg}
\affiliation{National Research Council of Canada, 100 Sussex Drive, Ottawa, Ontario K1N 5A2, Canada}
\affiliation{Department of Physics, University of Ottawa, 25 Templeton Street, Ottawa, Ontario, K1N 6N5 Canada}
\author{Khabat Heshami}
\affiliation{National Research Council of Canada, 100 Sussex Drive, Ottawa, Ontario K1N 5A2, Canada}
\affiliation{Department of Physics, University of Ottawa, 25 Templeton Street, Ottawa, Ontario, K1N 6N5 Canada}
\affiliation{Institute for Quantum Science and Technology, Department of Physics and Astronomy, University of Calgary, Alberta T2N 1N4, Canada}


\begin{abstract}
Non-Gaussian quantum states are crucial to fault-tolerant quantum computation with continuous-variable systems. Usually, generation of such states involves trade-offs between success probability and quality of the resultant state. For example, injecting squeezed light into a multimode interferometer and postselecting on certain patterns of photon-number outputs in all but one mode, a fundamentally probabilistic task, can herald the creation of cat states, Gottesman-Kitaev-Preskill (GKP) states, and more. We consider the addition of a non-Gaussian resource state, particularly single photons, to this configuration and show how it improves the qualities and generation probabilities of desired states. With only two modes, adding a single photon source improves GKP-state fidelity from 0.68 to 0.95 and adding a second then increases the success probability eightfold; for cat states with a fixed target fidelity, the probability of success can be improved by factors of up to four by adding single-photon sources. 
These demonstrate the usefulness of additional commonplace non-Gaussian resources for generating desirable states of light.
\end{abstract}

\maketitle


\section{Introduction}
The magic of quantum states is a mixed blessing: one rarely finds useful quantum states in nature. This makes clever state generation protocols essential to proposed quantum advantages from sensing to computation. In the particular context of light-based quantum systems, most of the known methods for generating non-Gaussian states, a prerequisite for photonic fault-tolerant quantum computation, have limited or trade-offs in success probabilities and state fidelities \cite{Peggetal1998,PhysRevLett.88.250401,Wengeretal2004,Zavattaetal2004,Ourjoumtsevetal2006,Vasconcelosetal2010,Bartleyetal2012,Etesseetal2015,BartleyWalmsley2015,Huetal2016,Birrittellaetal18,TakedaFurusawa2019,Eatonetal2019,Sabapathyetal2019,PhysRevA.100.052301,Quesadaetal2019}. We here explore how the combination of single-photon sources and Gaussian-boson-sampling (GBS) devices can improve creating these building blocks for photonic quantum information processing.

Light is a superior medium for encoding quantum information because of its speed and resistance to decoherence at room temperature, but the information is naturally encoded in the continuous variable (CV) degrees of freedom of harmonic oscillator modes as opposed to the discrete variable (DV) systems made from qubits, where most quantum computing algorithms and error correcting codes have been developed for the latter paradigm. At the start of the century, Gottesman, Kitaev, and Preskill introduced a method for robustly encoding a qubit in a harmonic oscillator mode~\cite{PhysRevA.64.012310} (requiring only one physical mode for one logical qubit \cite{TakedaFurusawa2019}) and their eponymous ``GKP states'' are now highly sought \cite{PhysRevA.97.022341,Tzitrinetal2021} for fast, decoherence-free, fault-tolerant quantum computation.

Another desirable set of CV states are ``cat states'' that are superpositions of two states that each on their own have highly classical properties \cite{Dodonovetal1974}. Cat states are useful for quantum information processing in their own right \cite{Cochraneetal1999,JeongKim2002,Ralphetal2003,Leghtasetal2013,Xu2023}, for breeding GKP states \cite{Vasconcelosetal2010,PhysRevA.97.022341,Konnoetal2023arxiv}, for quantum sensing protocols \cite{Zurek2001,Toscanoetal2006,Dalvitetal2006}, and for exemplifying the differences between quantum and classical theory \cite{Schrodinger1935cat,Souzaetal2008,Goldbergetal2020extremal,Akhtaretal2021}. We focus our attention on creating GKP and cat states, while our method can be broadly applied to generate any desired CV state.

Two checkpoints along the road to fault-tolerant quantum computation are boson sampling and its Gaussian (GBS) counterpart. Both inject nonclassical light into a linear optical network followed by photon counting, with output distributions that are challenging to recreate or sample from on a classical computer. The light injected for GBS is squeezed vacuum, a Gaussian state that is nonclassical in its ability to generate entanglement through a linear optical network \cite{GoldbergHeshami2021}, and this setup has been proposed for heralded generation of useful non-Gaussian states by choosing particular photon-number distributions among a subset of the output modes \cite{PhysRevA.100.052301,Quesadaetal2019}. A boson sampler, in contrast, directly injects what are arguably the most basic non-Gaussian resources, single-photon states. The classical hardness of predicting outcomes of quantum experiments was recently shown to depend on the amount of squeezing and number of single photons added to a network~\cite{Marshall:23}. Since the abilities to generate both squeezed vacuum and single photons are always improving, especially with promise of them both being feasible on similar physical platforms, we ask and answer the logical question: do single photons help GBS-like devices for generating interesting and useful states of light?

\section{Background}
\subsection{Non-Gaussian quantum states}

For a state defined on $N$ modes with bosonic annihilation operators $\hat{a}_i$, we define position and momentum operators as $\hat{q_i}=(\hat{a}_i+\hat{a}_i^\dagger)/\sqrt{2\hbar}$ and $\hat{p}_i=-i(\hat{a}_i-\hat{a}_i^\dagger)/\sqrt{2\hbar}$ and combine them into a single vector of operators $\bm{\hat{x}}=(\hat{q}_1,\hat{p}_1,\cdots,\hat{q}_N,\hat{p}_N)$. Gaussian states are uniquely characterized by the $2N$ first-order moments encoding the ``displacements'' of the state
$\bm{\xi}=\expval{\bm{\hat{x}}}$ and the $2N\times2N$ second-order moments encoded in the symmetric ``covariance matrix'' $\bm{V}$ with elements $V_{ij}=\expval{\frac{\hat{x}_i \hat{x}_j+\hat{x}_j \hat{x}_i}{2}}-\expval{\hat{x}_i}\expval{\hat{x}_j}$; whereas, non-Gaussian states need higher-order moments of $\bm{\hat{x}}$ to be uniquely specified.
Indeed, a convenient description of a state is through its Wigner function, which is responsible for the name ``Gaussian state'' because the latter have Gaussian Wigner functions~\cite{PRXQuantum.2.030204}:
\begin{equation}\label{eq:WignerGaussianState}
    W_G(\bm{x}) = \frac{\exp\left[\frac{1}{2}(\bm{x}-\bm{\xi})^\top\bm{V}^{-1}(\bm{x}-\bm{\xi})\right]}{(2\pi)^N\sqrt{\det\left(\bm{V}\right)}}.
\end{equation}

Due to the symmetrization in the covariance matrix, $\bm{V}$ does not have to be necessarily positive; instead, the standard uncertainty relations $[\hat{q}_i,\hat{p}_j]=i \hbar \delta_{ij}$ dictate that~\cite{PhysRevA.49.1567}
\begin{equation}\label{eq:CorrelationMatrix}
    \bm{V}+i\bm{\Omega}\geq 0,
\end{equation}
where $\Omega_{ij}=\delta_{i,j-1}-\delta_{i,j+1}$ are the elements of the $N$-mode symplectic form $\bm{\Omega}$. We will henceforth set $\hbar=1$.

Despite a wide range of tasks in quantum information can be accomplished using Gaussian states~\cite{RevModPhys.84.621, PhysRevA.66.032316, PhysRevA.96.062338}, non-Gaussianity has been recognized as an essential resource for quantum computation~\cite{LloydBraunstein1999,Bartlettetal2002,Ra2020-an,pmid:19995963,PhysRevLett.130.090602}. Non-Gaussian states can be particularly advantageous in achieving fault-tolerant universal quantum computing~\cite{PhysRevA.103.013710}. Additionally, non-Gaussian states exhibit greater entanglement, as measured by superadditive measures, compared to Gaussian states with the same correlation matrix~\cite{PRXQuantum.2.030204}.

The space of non-Gaussian states is indeed broad, with Schrödinger cat states and Gottesman-Kitaev-Preskill states being among the most well-known members of this group, alongside Fock states. Cat and GKP states have indeed several relevant applications~\cite{Asavanant:17, PRXQuantum.2.020101} that motivate the research conducted toward their more efficient generation~\cite{PhysRevA.105.063717, doi:10.1126/sciadv.abn1778, Eatonetal2019, PhysRevLett.128.170503}.

\subsubsection{GKP states}

GKP states are an encoding of qubits as two logical basis states $|\bar{0}\rangle$ and $|\bar{1}\rangle$ in a single harmonic oscillator mode that are simultaneous eigenstates of commuting displacements $e^{2\sqrt{\pi} iq_1 }$ and $e^{-2\sqrt{\pi}ip_1 }$ of momentum and position, respectively, where the displacement operators are termed stabilizers~\cite{PhysRevA.64.012310}. These states natively support error correction of amplitude- and phase-quadrature shifts because measurements of the stabilizers can detect such shifts without changing the state, making them ideal for fault-tolerant quantum information processing.

The encoded qubits can be set to be the following superposition of infinitely squeezed states
\begin{equation}
    \ket{\bar{0}_I}\equiv\sum_{n=-\infty}^{\infty}\ket{2n\sqrt{\pi}}_q; \quad \ket{\bar{1}_I}\equiv e^{-\sqrt{\pi}i p_1}\ket{\bar{0}_I},
\end{equation}
where the $I$ subscripts denote the ideal states and the $q$ subscript denotes a position eigenstate. After verifying that these states indeed have the correct eigenvalue properties, one can observe the following two displacement operators to be the encoded Pauli operators acting on the encoded qubit states:
\begin{equation}
\bar{Z}=\exp\left(i\sqrt{\pi}q\right), \quad \bar{X}=\exp\left(-i\sqrt{\pi}p\right).
\end{equation} 
These satisfy the commutation relation
\begin{equation}
    \bar{X}\bar{Z}=-\bar{Z}\bar{X}
\end{equation} which allows them to detect shifts along $p_1$ and $q_1$ up to $\sqrt{\pi}/2$. That the stabilizers are the squares of the logical operators $\bar{X}$ and $\bar{Z}$ accords with the squares of Pauli matrices being identity.

The Wigner function of the state $\ket{\bar{0}_I}$ is

\begin{equation}\label{WignerFunction}
        W_{\mathrm{GKP}\, \bar{0}_I}(q,p) = \sum_{s,t = -\infty}^{\infty} (-1)^{st}\delta\left(p-\frac{s\sqrt{\pi}}{2}\right)\delta\left(q-t\sqrt{\pi}\right).
\end{equation}

Since the ideal state defined here is nonnormalizable and made of infinitely squeezed states, we can work with normalizable states that approximate the ideal ones. There exist several possible approximations for GKP states, and the canonical approximation \cite{PhysRevA.101.032315} is one of the most common ones. In the canonical approximation, the Dirac deltas in the Wigner functions are approximated by Gaussian functions with finite standard deviation $\Delta$, and the overall Wigner function is enveloped in another Gaussian function with width $\kappa^{-1}$
\begin{equation}\label{eq:CanonicalNormalization}
    \ket{\bar{0}_I}\rightarrow\ket{\bar{0}_{\Delta,\kappa}}\propto\sum_{n=-\infty}^{\infty}e^{\frac{1}{2}\kappa^2(2n\sqrt{\pi})^2}\bar{X}^2\ket{\Delta}_q,
\end{equation}
where
\begin{equation}
    \braket{q}{\Delta}=\left(\frac{1}{\pi\Delta^2}\right)^\frac{1}{4}e^{-\frac{q^2}{2\Delta^2}}.
\end{equation}

\begin{figure}[htbp]
\begin{center}
    \includegraphics[trim=30 200 100 100, clip,width=0.5\textwidth]{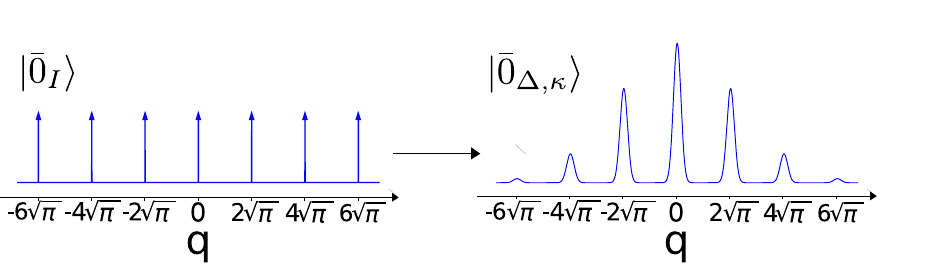}
\caption{Distribution of the position $q$ for the ideal GKP state $\ket{\bar{0}_I}$ and the normalized GKP state $\ket{\bar{0}_{\Delta,\kappa}}$ with $\Delta=\kappa=1/4$.}
\label{fig:ApproximateGKP}
\end{center}
\end{figure}

The ideal GKP state is then restored when $\Delta,\kappa\rightarrow0$. The Wigner functions, integrated over $p$, of the ideal basis states $\ket{\bar{0}}_{\text{GKP}}$ and its approximation $\ket{\Tilde{0}}_{\text{GKP}}\equiv \ket{\bar{0}_{\Delta,\kappa}}$ are shown in Fig.~\ref{fig:ApproximateGKP}.  An extremely useful property of GKP states is that Gaussian noise from effects such as finite squeezing can be converted into errors on the encoded qubit, which can in turn be corrected by error-correcting codes known for DV quantum computation \cite{Larsenetal2021}.

\subsubsection{Cat states}

Cat states are symmetric (even) or antisymmetric (odd) pure superpositions of two coherent state with opposite amplitudes $\pm\alpha$
\begin{equation}\label{eq:catstatedefinition}
    \ket{\text{cat}_{\mathrm{e/o}}(\alpha)} = \frac{1}{N_\pm(\alpha)}\left(\ket{\alpha}\pm\ket{-\alpha}\right),
\end{equation}
where $N_{\pm}(\alpha)=\sqrt{2(1\pm e^{-2\abs{\alpha}^2})}$.
The cat state is a non-Gaussian state with Wigner function given by~\cite{10.1093/acprof:oso/9780198509141.001.0001}

\begin{multline}
    W_{\text{cat}_{\mathrm{e/o}(\beta)}}\left(\alpha\right)=\frac{2}{\pi (1\pm e^{-2\abs{\beta}^2})}\\ \left(e^{-2\abs{\alpha-\beta}^2}+e^{-2\abs{\alpha+\beta}^2}\pm2e^{-2\abs{\alpha}^2}\cos(4\Im(\alpha\beta^*))\right).
\end{multline}

By expanding the coherent states in terms of Fock states, we observe that the even (odd) cat states can also be expressed as a linear superposition of solely even (odd) photon-number states
\begin{gather}
    \ket{\text{cat}_{\mathrm{e}}(\alpha))}\propto\frac{\alpha^0}{\sqrt{0!}}\ket{0}+\frac{\alpha^2}{\sqrt{2!}}\ket{2}+\frac{\alpha^4}{\sqrt{4!}}\ket{4}+\dots;\label{eq:catFockExpansioneven}\\
    \ket{\text{cat}_{\mathrm{o}}(\alpha))}\propto \frac{\alpha^1}{\sqrt{1!}}\ket{1}+\frac{\alpha^3}{\sqrt{3!}}\ket{3}+\frac{\alpha^5}{\sqrt{5!}}\ket{5}+\dots.\label{eq:catFockExpansionodd}
\end{gather}

Cat states are extremely versatile, with applications in quantum communication and quantum metrology~\cite{Gaidash_2019, PhysRevA.100.032318}. Encoding qubits in cat states can help fault-tolerant quantum computation~\cite{Cochraneetal1999, PhysRevResearch.4.043065, MIRRAHIMI2016778}. They can be employed also as a resource to prepare GKP grid states~\cite{PhysRevA.97.022341, Vasconcelosetal2010}, or can be used themselves to perform quantum error correction~\cite{PhysRevLett.77.793, PhysRevA.106.022431} and have been generated using nonlinear media~\cite{He2023}.

\subsection{Non-Gaussian optical state preparation}
Any $N$-mode Gaussian state can be generated by applying Gaussian operators to a given initial Gaussian state, e.g. the $N$-mode vacuum state $\ket{0}^{\otimes N}$. Gaussian operations are those operations defined by unitary evolutions under Hamiltonians that are at most quadratic in  creation and annihilation operators $\hat{a}^\dag, \hat{a}$.
Under these evolutions, the displacement vector $\bm{\xi}$ and covariance matrix $\bm{V}$ that univocally determine a Gaussian state are transformed according to the so-called symplectic transformations~\cite{Serafini2017-gp}
\begin{equation}
    \bm{V} \rightarrow \bm{FVF}^\top \quad \text{and} \quad \bm{\xi} \rightarrow \bm{F\xi} + \bm{d},
\end{equation}
where $\bm{d}$ is the $2N$ real vector of displacements, and $\bm{F}$ satisfies
\begin{equation}
    \bm{F\Omega F}^\top = \bm{\Omega}.
\end{equation}
In this way, the transformed $\bm{\xi}$ and $\bm{V}$ respectively remain a $2N$-dimensional real vector and a $2N\times2N$ real symmetric matrix that satisfies Eq.~\eqref{eq:CorrelationMatrix}. These transformed parameters unambiguously define a new Gaussian state.

From a practical perspective, in optical implementations, all Gaussian transformations can be realized using squeezing and displacement operators, as well as an $N$-mode interferometer composed of beam splitters and phase shifters. These operations are commonly implemented in current experimental setups, with the squeezing amplitude being the main constraining parameter.

The actions of a squeezer $S$, and beam splitter $B$ are respectively defined in terms of creation $\hat{a}^\dagger_i$ and annihilation operator $\hat{a}_i $ on the $i$th mode as
\begin{gather}
     S(r) = \exp\left(\frac{r}{2}\left(\hat{a}^2-\hat{a}^{\dagger2} \right)\right),\\
       B_{ij}(\theta,\varphi) = \exp\left(\theta\left(e^{i\varphi}\hat{a}_i\hat{a}^\dagger_j-e^{-i\varphi}\hat{a}^\dagger_i\hat{a}_j\right)\right).\label{eq:BeamSplitter}
\end{gather}

Non-Gaussian unitary evolutions are generated by Hamiltonians containing terms that are at least cubic in the creation and annihilation operators~\cite{Brauer:21}. However, these operators are difficult to realize on optical tables~\cite{PhysRevA.91.032321,PhysRevA.107.022422}. An alternative and often preferred approach to generating non-Gaussian states involves performing a measurement on a Gaussian state $\rho$~\cite{Genoni_2010, Davis_2021} such that
\begin{equation}
    \rho\rightarrow\frac{\sum_jX_j\rho X_j^\dagger}{\Tr{\rho\sum_j X_j^\dagger X_j}},
\end{equation}
where $X_j$ represents a linear operator in the Fock space. For instance, in the case of photon-number resolving detectors (PNRDs), we have $X_{ijk\dots}=\ketbra{ijk\dots}{ijk\dots}$. Performing a measurement on $m$ modes of an $n$-mode Gaussian state using PNRDs provides a practical method for conditionally generating a $(n-m)$-mode non-Gaussian state. The scheme originally developed for Gaussian boson sampling can indeed be modified for this purpose.

\subsection{Gaussian boson sampling and state generation}\label{sec:StateGenwGBS}
Gaussian boson sampling is a quantum computational method that provides a significant advantage over classical computers for some specific problems~\cite{PhysRevLett.119.170501}. The method consists of two steps. First a $n$-mode Gaussian state is generated by injecting $n$ squeezed vacuum states into a linear interferometer. Then, each mode of the output Gaussian state is measured with a PNRD. The measurement pattern is the output of the computation. Finding the probability of any measurement pattern is in fact a problem with several relevant applications but classically hard to solve~\cite{VALIANT1979189}. Since it is only possible to sample from such an output distribution, the applications are tempered, but even the sampling problem is computationally challenging for classical machines.

\begin{figure}[htbp]
\begin{center}
    \includegraphics[trim=20 110 40 50, clip,width=0.5\textwidth]{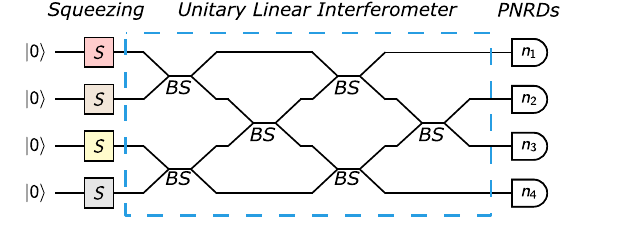}
\caption{GBS schematic with four modes. The $N(N-1)/2$ beam splitters and phase shifters (BS) are arranged in the so-called rectangular scheme~\cite{Clements:16}. $BS$ label the beam splitters whose action in terms of the creation (annihilation) operators $a^\dagger_i$ ($a_i$) of the $i$th mode depend on the angles $\theta$ and $\varphi$ as defined in Eq.~\eqref{eq:BeamSplitter}.}
\label{fig:GBS4modes}    
\end{center}
\end{figure}

\begin{figure}[htbp]
\begin{center}
    \includegraphics[trim=90 90 90 60, clip,width=0.4\textwidth]{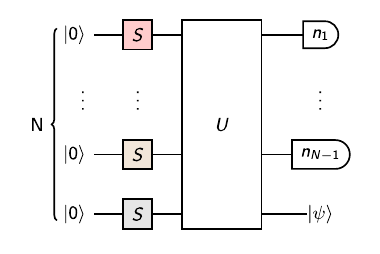}
\caption{GBS device with $n$ squeezed displaced input vacuum states and $n-1$ PNRDs.}
\label{fig:GBSdevicepure}
\end{center}
\end{figure}

The GBS device described thus far, and displayed in Fig.~\ref{fig:GBS4modes} for four modes, 
can be slightly modified to serve as a conditional source of non-Gaussian states. In fact, if some modes are left unmeasured, the corresponding heralded mode is proven to be non-Gaussian, provided that at least one photon is detected in the other modes.
In general, in an $n$-mode GBS-like device, $(n-m)$ modes are measured to produce an $m$-mode non-Gaussian state. However, from now on, we will consider the case where only one mode is left unmeasured because our goal is to produce single-mode non-Gaussian states.
An example of GBS device used for state preparation~\cite{PhysRevA.101.032315, Takase2023}, with $n$ input modes, and $n-1$ PNRDs is shown in Fig.~\ref{fig:GBSdevicepure}. Any unitary operation $U(\bm{\theta})$ on the $n$ modes can be realized using a linear interferometer with as few as $n(n-1)/2$  beam splitters and phase shifters~\cite{PhysRevLett.73.58, Clements:16}.

When no photons are detected by the PNRDs the Wigner function of the heralded state is Gaussian and is unambiguously determined by squeezing and displacement coefficients.
In general, if $n_T$ photons are detected in the measurement, then the output state can be expressed as a squeezed and displaced linear superposition of Fock states

\begin{equation}\label{eq:Stellar}
    \ket{\psi_{\text{out}}}=\hat{D}(d)\hat{S}(\zeta)\sum_{n=0}^{n_{\text{max}}}c_n\ket{n}.
\end{equation}

Eq.~\eqref{eq:Stellar} is known as the stellar representation of the state $\ket{\psi}$ with a stellar rank $r^* = n_\text{max}\leq n_T$~\cite{PhysRevLett.124.063605}. The value of $n_\text{max}$ depends on the connections established between the modes by $U(\bm{\theta})$. Gaussian states indeed have a stellar rank of $r^*=0$, while $n_{\text{max}}$ is equal to $n_T$ only when the heralded mode is fully connected with all the detected modes.
Furthermore, the number of independent coefficients $c_n$ is constrained by the angles $\bm{\theta}$ that unambiguously determine the $N$-mode linear interferometer. In fact, the maximum number of independent parameters is given by $\mathcal{D}=(N+2)(N-1)/2$.

\section{GBS-like device with Single Photon sources}

Conditional state generation via a GBS-like device uses measurements with PNRDs as a source of non-Gaussianity, which is required to generate non-Gaussian states. The nonlinearity of the measurement is, in fact, a technique frequently employed to produce quantum optical non-Gaussian states~\cite{PhysRevA.72.033822,Dakna1998,PhysRevLett.107.163602,PhysRevLett.114.170802,winnel2023deterministic,Melalkia_2023}. It has even been used to develop a systematic technique that generates GKP qubits by exploiting the symmetries of the target states~\cite{Takase2023}. Conditional state generation provides an alternative to the use of challenging-to-realize deterministic non-Gaussian operators and offers easier implementation. However, it has a drawback due to the probabilistic nature of the outcome.
The desired state is actually generated only when certain conditions are met. If the conditions are not satisfied, one has two options: either choose to encode the qubit into a Gaussian state, thereby losing protection against noise, or run the non-Gaussian source again until the conditions are fulfilled. Each iteration of the source, however, delays the encoding and actual processing of the information, exposing the system to decoherence and photon loss. For this reason, enhancing the probability of non-Gaussian state generation is crucial for scalable quantum computation. Multiplexing can be used to get around this problem but at the cost of added loss in the system.
As an alternative, to increase the probability of success, we can introduce an additional source of non-Gaussianity alongside the measurements with PNRDs. Specifically, instead of using vacuum states, non-Gaussian states can be employed in some of the modes of the interferometer. The generation of single-mode non-Gaussian states has indeed been theorized \cite{Peggetal1998,PhysRevA.59.1658,PhysRevA.82.053812,PhysRevA.64.063818,PhysRevA.62.013802,PhysRevA.63.055801,Miranowicz_2005,Eatonetal2019,Mattos:23} and successfully realized \cite{PhysRevLett.88.250401,PhysRevLett.88.113601,Bimbardetal2010} in less general setups by incorporating non-Gaussian resources at the input and utilizing PNRDs.

Naturally, this approach is helpful as long as the elected input non-Gaussian states can be produced with a highly efficient alternative approach. Single photon states, for example, can be generated with several efficient protocols~\cite{Lounis_2005,doi:10.1126/sciadv.abc8268} that are compatible \cite{Maringetal2023arxiv,Larocqueetal2023arxiv} with the integrated photonics platforms on which GBS experiments have been demonstrated. In fact, squeezed single-photon states proved to be advantageous compared to squeezed vacuum in other applications~\cite{Olivares_2016, PhysRevA.78.052304} and practical routes to squeezing single photons have been demonstrated \cite{PhysRevLett.113.013601,Engelkemeieretal2020,Engelkemeieretal2021arxiv}, following the general trend of developing the ability to add light to non-vacuum light fields \cite{Zavattaetal2004,Zavattaetal2005,Parigietal2007}.
In this work, we evaluate the impact of using single photon Fock states instead of vacuum states on the performance of non-Gaussian state generation. We compare the results achieved with a standard GBS-like device depicted in Fig.~\ref{fig:GBSdevicepure} and thoroughly described in~\cite{PhysRevA.100.052301} where the only non-Gaussian resource is the type of measurement being conducted. The two schemes are essentially the same when no single-photon sources are integrated into the protocol. Furthermore, an equivalence can be established between the two schemes even when single-photon Fock states are used as inputs in our approach. This equivalence arises when, in the full Gaussian scheme, one considers a portion of the interferometer as serving as a conditional single-photon source, achieved through generalized photon subtraction at the cost of one additional ancillary mode per input Fock state.
We observe that the theory developed for GBS enables the analytical determination of the generated heralded state when $(n-m)$ modes are measured out of an $n$-mode linear interferometer receiving squeezed vacuum states as input. However, for a desired target state, finding the parameters that ensure high levels of fidelity and probability in the measurement pattern becomes impractical to solve analytically. In such cases, numerical optimizations are often preferred to address this problem effectively.
The libraries provided by Xanadu are optimal tools for simulating GBS devices and optimizing their tunable parameters. Specifically, the library Mr Mustard is employed here both for GKP and Cat states. 

The approximate GKP state used as a target is the truncated core state of 
\begin{equation}\label{eq:GKP4}
   \ket{0_{A4}}=S(r)\underbrace{\sum_{n=0}^4 c_n\ket{n}}_{\text{Core state}}
\end{equation}
where $\ket{0_{A4}}$  maximizes $\expval{0_\Delta}{0_{A4}}$, where $\ket{0_\Delta}$ corresponds to the canonical normalization of the ideal GKP state $\ket{0_I}\rightarrow\ket{0_{\Delta,\kappa}}$ defined in Eq.~\ref{eq:CanonicalNormalization} with $\kappa=\Delta$. In fact, the same state is used as a target by Tzitrin \textit{et al.} in~\cite{PhysRevA.101.032315}. 

The target cat state is instead, in turn, defined as  
\begin{equation} 
 \ket{\text{cat}_e(2)}=\frac{1}{\sqrt{2}}(\ket{2}+\ket{-2}),
\end{equation} 
that is the even cat state defined in Eq.~\eqref{eq:catstatedefinition} with amplitude $\alpha=2$.

Since both the probability $p$ and the fidelity $\mathcal{F}$ to the target state are relevant features of a desirable source, the reward function maximized by the optimized parameters is set as a linear combination of these figures of merit. While the choice of the reward function is essentially arbitrary, it requires careful consideration because of the nature of the optimizer used in simulations, especially when dealing with larger circuits. Prioritizing probability over fidelity runs the risk of generating a state far from the target. Conversely, assigning excessive weight to fidelity can slow down optimization and lead to the generation of nearly unattainable states.

For our specific objective, we opt for a reward function $\mathcal{F}+p$ that evenly balances both probability and fidelity in all the optimizations, regardless of the number of modes and the target. This is simply a means to an end: we then report the results of all of these optimizations in terms of both $\mathcal{F}$ and $p$ to see how single-photon sources can simultaneously improve both quantities. The actual relationship between the number of single-photon sources and a combined figure-of-merit parameter is more complicated; \textit{quod vide} Fig.~\ref{fig:ResultsvsSPSs3modes}.

\begin{figure}[htbp]
\begin{center}
    \includegraphics[trim=15 50 70 70, clip,width=0.5\textwidth]{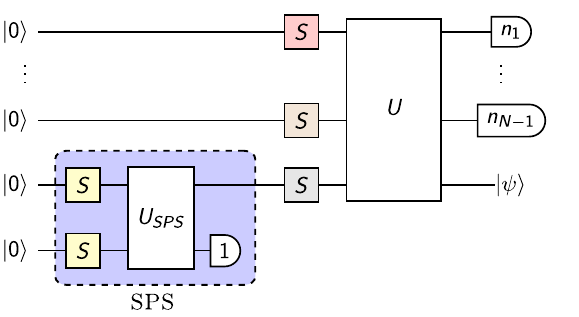}
\caption{GBS-like device with one single photon source in the purple box.}
\label{fig:GBSwSPSscheme}
\end{center}
\end{figure}

The squeezing amplitude $r$ is a tunable parameter optimized in the simulations, but a maximum energy threshold is set for it at 12dB.
As a final remark, ancillary GBS has been introduced as a single photon source instead of directly using the single photon Fock state $\ket{1}$ in the input modes of the interferometer. The example of a scheme of a GBS-like device with one single photon source into the $N$-th mode is shown in Fig.~\ref{fig:GBSwSPSscheme}.
This choice is made because non-Gaussian input states could potentially make the simulation with the libraries less reliable (see Appendix~\ref{appendix:SPS} for a discussion of how we numerically introduce single-photon sources based on physical architectures). 

The photon subtraction technique is also used to prepare squeezed single-photon states~\cite{Olivares_2005}. Here, instead, the simulations of the inline squeezing of single photons are performed independently of the state generation process since optimal single-photon sources are intended to replace the GBS-like devices employed in the simulation. In fact, we assume to have a deterministic single-photon source.

The measurement patterns are chosen with a small total number of subtracted photons to prioritize the probability of generating the state over the fidelity with the target state, which is already satisfactory for many applications. We indeed expect to increase the fidelity at the cost of smaller probability when more photons are detected~\cite{Sabapathyetal2019}. This choice also helps reduce the run time of the optimizations. We investigate multiple measurement patterns to evaluate the expected trend of the results with the number of detected photons. In all measurements, however, all the PNRDs detect at least one photon. 

Futhermore, we observe that both the target GKP and even cat states can be expressed as linear combinations of even number states in the Fock representation. The same argument holds for the squeezed vacuum states in each input mode 

\begin{equation}
    S(r)\ket{0}=\sum_{n=0}^\infty\frac{\tanh^n r}{\sqrt{\cosh r}}\frac{\sqrt{(2n)!}}{2^nn!}\ket{2n}.
\end{equation}
 
Since beam splitters and phase shifters conserve the number of photons in the state, the $N$-mode output state can be expressed as 

\begin{equation}
    \ket{\psi_{\text{out}}}=\sum_{n=0}^\infty \sum_{m_1+\dots+m_M=2n} c_{m_1,\dots m_N}\ket{m_1,\dots m_N}.
\end{equation}
 
As a consequence, in order to herald a state with an even number of photons, the PNRDs have to detect an even number of photons in total. 
If, instead, single-photon sources are used, then the necessary parity of the measured photons depends on how many squeezed single-photon states are present in the input.

\section{Results and discussion}

\begin{figure}[htbp]
\begin{center}
    \includegraphics[width=0.45\textwidth]{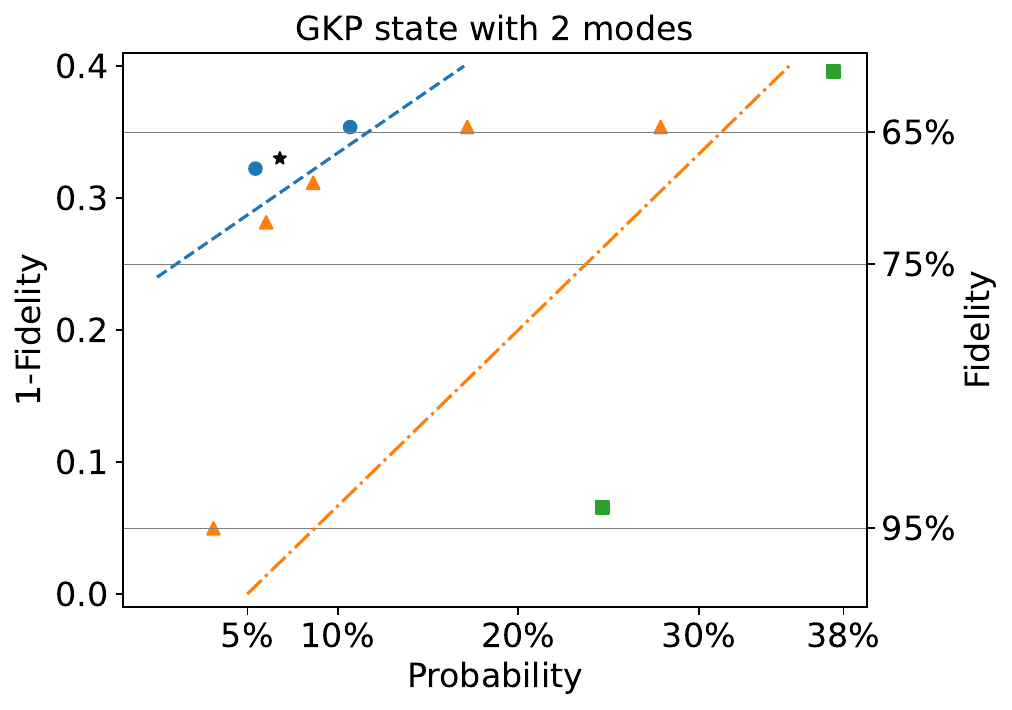}
\caption{Plot of the results obtained with a 2-mode GBS-like device when the target state is the approximate GKP state $\ket{0_{A4}}$. The blue dots represent the optima found without single photon sources. The star refers to the best result reported in~\cite{PhysRevA.101.032315} without any single photon sources ($n=0$). The orange triangles depict the results obtained with one single photon source. The green squares correspond to the results obtained with two single photon sources. Diagonal lines limit regions in the fidelity-probability space where results of the corresponding colour can be found.}
\label{fig:ResultsGKP}
\end{center}
\end{figure}

First, we consider the case of a 2-mode GBS-like device and target state given by the approximate GKP states $\ket{0_{A4}}$ defined in Eq.~\ref{eq:GKP4}.
The probabilities $p$ and fidelities $\mathcal{F}$ obtained with $n=0, 1 $ and 2 single photon sources are plotted in Fig.~\ref{fig:ResultsGKP}. The best results are also reported in Table~\ref{tab:GKPResults}.
\begin{table}[h!]
\begin{center}
\begin{tabular}{ | c | c | c | c | } 
 \hline
 $n$ & 1-$\mathcal{F}$ & $p$ & $n_T$  \\ 
  \hline
 \multirow{2}{0.5em}{0} & 0.35 & $11\%$ & 4 \\
 & 0.32 & $5.4\%$ & 4\\ 
\hline
\multirow{3}{0.5em}{1} & 0.35 & $27\%$ & 3\\
 & 0.31 & $8.6\%$ & 3\\ 
  & $4.9\times10^{-2}$ &$3.1\%$ & 5\\ 
\hline
\multirow{2}{0.5em}{2} & 0.40 & $37\%$ & 2\\
 & $6.6\times10^{-2}$ & $24\%$ & 2\\ 
\hline
\end{tabular}
\caption{Table showing the fidelities $\mathcal{F}$ and probabilities $p$ obtained by a 2-mode GBS-like device when the target state is the approximate GKP state $\ket{0_{A4}}$. The parameter $n$ represents the number of input modes receiving the single-photon state $\ket{1}$, while $n_T$ denotes the number of photons observed in the detected mode.}
\label{tab:GKPResults}

\end{center}
\end{table}

\begin{figure}[htbp]
\begin{center}
    \includegraphics[width=0.5\textwidth]{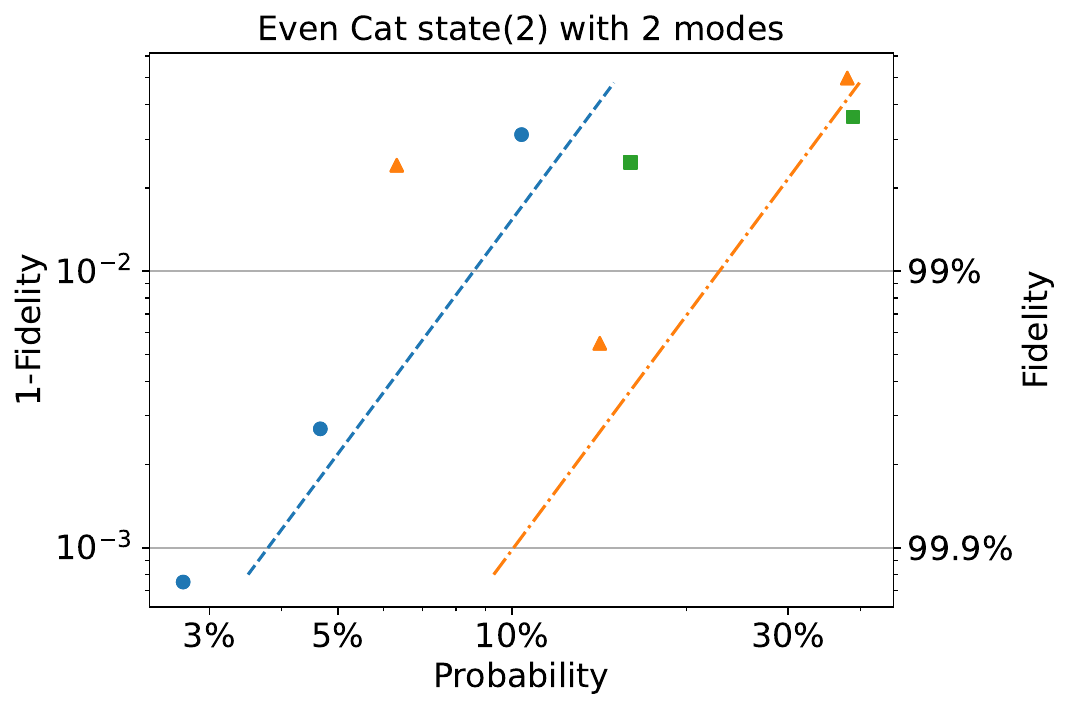}
\caption{Plot of the results obtained with a 2-mode GBS-like device when the target state is the cat state $\ket{\text{cat}_e(2)}$. The blue dots refer to the optima found without single photon sources. The orange triangles correspond to the results obtained with one single photon source. The green squares indicate the results obtained with two single photon sources.}
\label{fig:ResultsCAT2}
\end{center}
\end{figure}

\begin{figure}[htbp]
\begin{center}
    \includegraphics[width=0.5\textwidth]{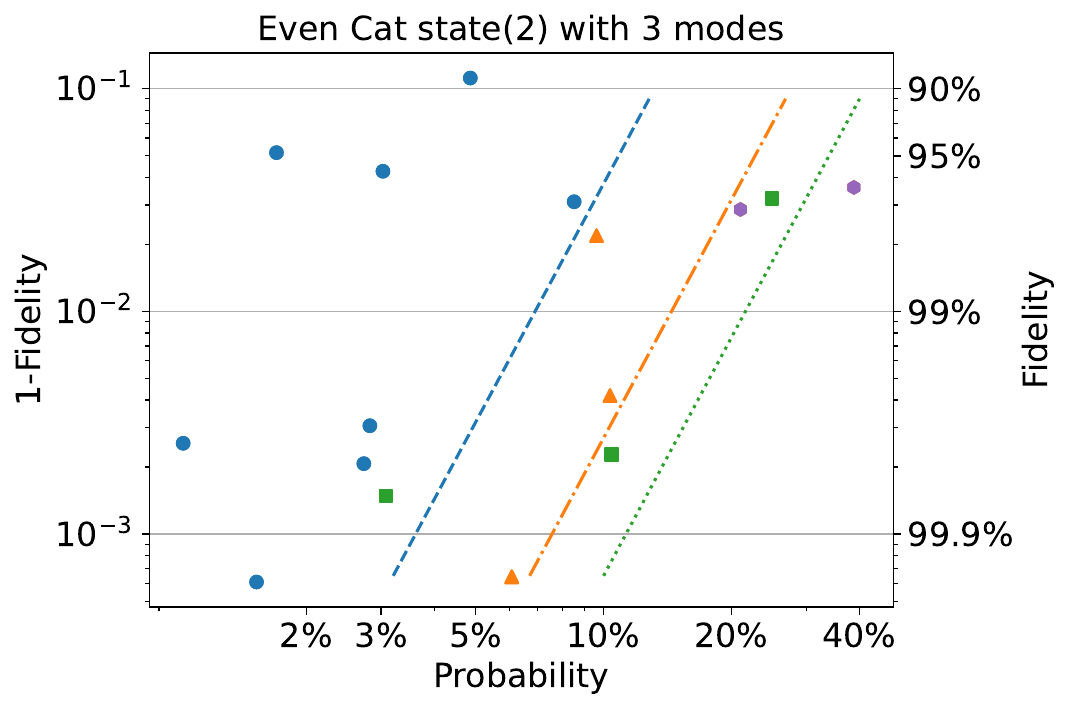}
\caption{Plot of the results obtained with a 3-mode GBS-like device when the target state is the cat state $\ket{\text{cat}_e(2)}$. The blue dots refer to the optima found without single photon sources. The orange triangles correspond to the results obtained with one single photon source. The green squares indicate the results obtained with two single photon sources. The purple hexagons indicate the results obtained with three single photon sources. Diagonal lines limit regions in the fidelity-probability space where results of the corresponding colour can be found.}
\label{fig:ResultsCAT3}
\end{center}
\end{figure}

The results obtained with the target state given by the even cat state of amplitude $\alpha=2$ are displayed in Fig.~\ref{fig:ResultsCAT2} and Fig.~\ref{fig:ResultsCAT3} , for two-mode and three-mode GBS-like devices, respectively. The best results are listed in Table~\ref{tab:CATResults2} and Table~\ref{tab:CATResults3}.

\begin{table}[]
\begin{center}

\begin{tabular}{ | c | c | c | c | } 
 \hline
 $n$ & 1-$\mathcal{F}$ & $p$ & $n_T$  \\ 
  \hline
 \multirow{3}{0.5em}{0} & 3.1$\times10^{-2}$ & $10\%$ & 2 \\
 & 2.7$\times10^{-3}$ & $4.7\%$ & 4\\ 
  & 7.5$\times10^{-4}$ & $2.7\%$ & 6 \\ 
\hline
\multirow{2}{0.5em}{1} & 5.0$\times10^{-2}$ & $38\%$ & 1\\
 & 5.5$\times10^{-3}$ & $14\%$ & 3\\ 
\hline
\multirow{1}{0.5em}{2} & 3.6$\times10^{-2}$ & $39\%$ & 2\\ 
\hline

\end{tabular}
\caption{Same as Table~\ref{tab:GKPResults} but with the target state being the even cat state of amplitude $\alpha=2$.
}
\label{tab:CATResults2}

\end{center}
\end{table}

\begin{table}[]
\begin{center}

\begin{tabular}{ | c | c | c | c | } 
 \hline
 $n$ & 1-$\mathcal{F}$ & $p$ & $n_T$  \\ 
  \hline
 \multirow{4}{0.5em}{0} & 3.1$\times10^{-2}$ & $8.5\%$ & 2 \\
 & 2.1$\times10^{-3}$ & $2.7\%$ & 4\\ 
 & 3.1$\times10^{-3}$ & $2.8\%$ & 4\\ 
  & 6.1$\times10^{-4}$ & $1.5\%$ & 6 \\ 
\hline
\multirow{2}{0.5em}{1} & 4.2$\times10^{-3}$ & $10\%$ & 3\\
 & 6.4$\times10^{-4}$ & $6.1\%$ & 5\\ 
\hline
\multirow{2}{0.5em}{2} & 3.2$\times10^{-2}$ & $25\%$ & 2\\ 
 & 2.3$\times10^{-3}$ & $10\%$ & 4\\ 
\hline
\multirow{1}{0.5em}{3} & 3.6$\times10^{-2}$ & $39\%$ & 3\\ 
\hline

\end{tabular}

\caption{Same as Table~\ref{tab:CATResults2} but with 3-mode GBS-like device.
}
\label{tab:CATResults3}

\end{center}
\end{table}

As expected, the results reported in Table~\ref{tab:CATResults2} for the case without any single photon sources align with those presented in Table II of~\cite{PhysRevA.100.052301}, where slightly improved fidelity and probability are achieved only with squeezing amplitudes exceeding the threshold considered in our work.

\begin{figure}[htbp]
\begin{center}
    \includegraphics[width=0.5\textwidth]{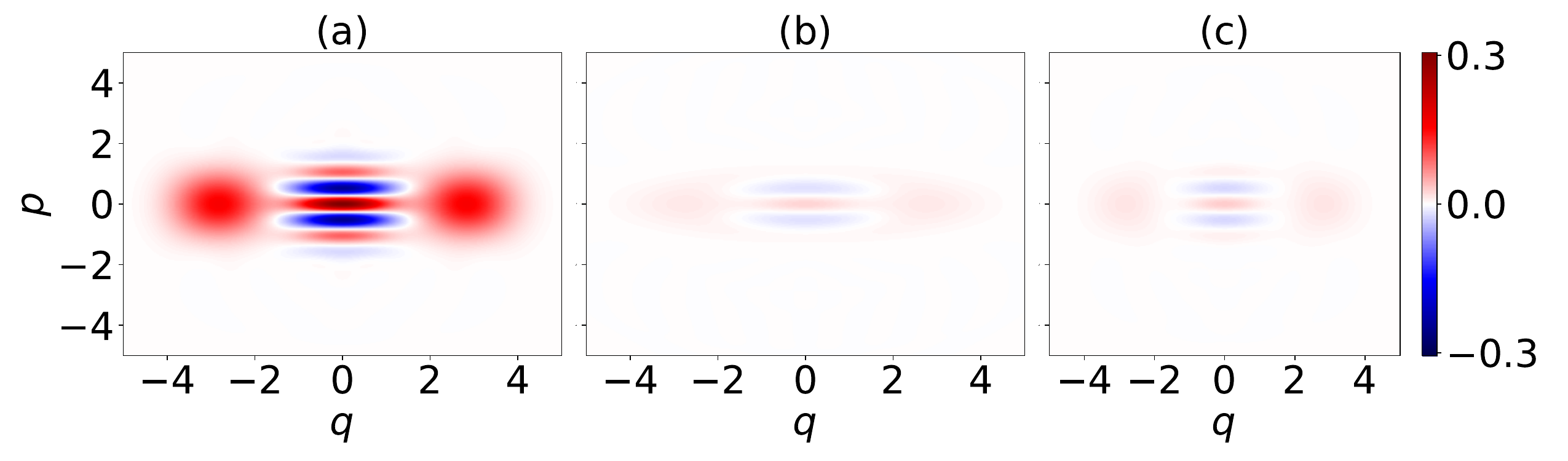}
\caption{Colour maps of the Wigner functions of  (a) the target cat state, (b) the state generated using a GBS-like device without any single photon sources, and (c) the state generated using a GBS-like device with one single photon source. The probability of generating the state $P$ is higher when a single Fock state is injected into an input mode, $P = 10.5\%$. In comparison, the probability achieved solely with Gaussian inputs is $P = 8.5\%$. Moreover, an improvement in fidelity is observed, with an increase from $96.9\%$ to $99.6\%$. This enhancement is visually apparent in the Wigner function plotted. 
The differences in colour shading between the generated states and the target state correspond to reduced values of the Wigner function, where the norm of this function corresponds to the probability of generating the displayed states.}
\label{fig:WingerColourMaps}
\end{center}
\end{figure}

Each data point depicted in the two-dimensional plots corresponds to an optimization run utilizing the Python libraries ``Mr Mustard" developed by Xanadu. 
Mr Mustard ~\cite{XanaduAI} is a Python library particularly suited for simulating Gaussian boson sampling devices. It efficiently handles the phase space representation of Gaussian states and Gaussian channels, and it possesses several built-in functions simulating the action of various optical devices used in GBS protocols, such as beam splitters, phase shifters, squeezers, Mach-Zender interferometers, and PNRDs. The description in the number representation, especially useful at the last measurement stage of the protocol, can be realized with this library with arbitrary precision and cutoff for any Gaussian state. Finally, the library is equipped with optimization routines expressly dedicated to the specific gate. Unitary optimization is used in our optimizations for the linear interferometer, while optimization in Euclidean space is adopted for the squeezing amplitudes at the input modes.
On the vertical axis, the plots report the fidelity between the state generated by simulating a photonic circuit with the given optimized parameters and the predefined target state. On the horizontal axis, it is reported the probability that the PNRDs measure the detection pattern that guarantees the generation of the desired state.  
Irrespective of the number of Fock states in the input modes of the interferometer, the optimization of the circuit's tunable parameters can yield varying outcomes. These outcomes depend on the initial parameters and the specific measurement pattern considered. As a consequence, the optimizations over a GBS-like device with a fixed number of modes and of input Fock states can lead to possibly very different results in terms of fidelity and probabilty. Nevertheless, a general trend can be identified, and it is highlighted by straight lines in the fidelity-probability space that demarcate the regions where given classes of results are present and where they are absent. 

As expected~\cite{Sabapathyetal2019}, at a fixed number of single photon sources, the fidelity increases at the expense of lower probability when more photons are measured. 
Overall, we observe that the efficiency of the non-Gaussian state source, realized with a boson sampling-like device, experiences improvement when single-photon states are injected into the input modes of the interferometer. For example, when considering the three-mode GBS-like device, the probability of generating a cat state with fidelity around $97\%$ steadily increases, potentially doubling its value, every time a squeezed single-photon state is used instead of a squeezed vacuum. 
The improvement in the state generation is qualitatively displayed in Fig.~\ref{fig:WingerColourMaps} through a comparison of the Wigner functions of the target cat state and the states generated using a three-mode GBS-like device with and without SPSs.

From a comparison between the outcomes related to cat-state generation with 2- and 3-mode networks, no improvements emerge from using more modes. Indeed, it appears that the additional resources do not play a significant role in increasing fidelity with the target state, as one might expect. In fact, in the context of state generation with purely Gaussian states, the number $\mathcal{D}$ of independent parameters $c_i$ in~\eqref{eq:Stellar} grows quadratically with the number of modes, as conjectured in~\cite{PhysRevA.100.052301}. However, if the total number of measured photons $n_T<\mathcal{D}$, the problem of finding the best parameters is underdetermined, and the optimal result can be found with smaller Gaussian states. A similar argument holds even when the number of measured photons $n_t\geq\mathcal{D}$, but the target state can be well approximated by a quantum state with a small rank, as is the case for the cat state with amplitude $\alpha=2$. In this scenario, the state that can be generated by the smaller circuit exhibits satisfactory fidelity, and the search for a better state attainable with additional modes becomes unfeasible for the optimizer.

\begin{figure}[htbp]
\begin{center}
    \includegraphics[width=0.45\textwidth]{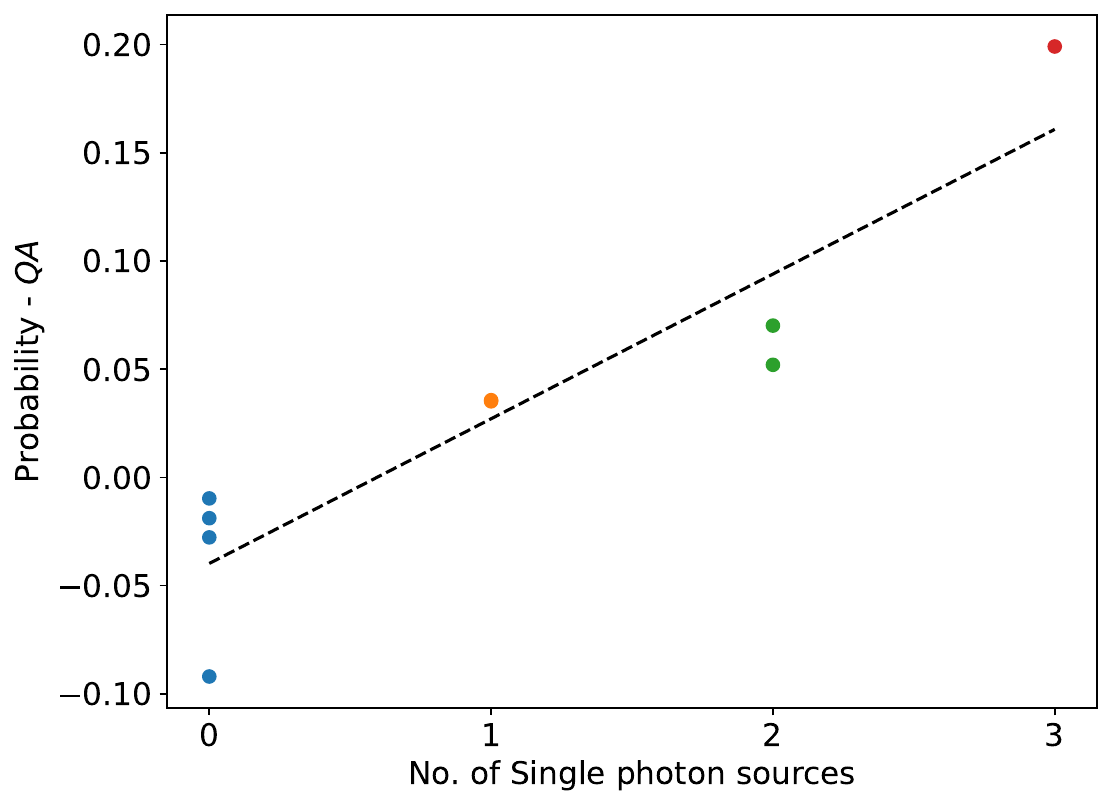}
\caption{Plot depicting the difference between the probability of state generation and the quantum angle with the target state as a function of the number of single photon sources, using the data from the Table~\ref{tab:CATResults3}. The dashed line represents the linear regression of the data, with a slope of approximately $6.7 \times10^{-2}$ and an intercept of approximately -$4.0 \times10^{-2}$; note that this difference can indeed be negative}.
\label{fig:ResultsvsSPSs3modes}
\end{center}
\end{figure}

Finally, in Fig.~\ref{fig:ResultsvsSPSs3modes}, we plot the results presented in Table III, by showing how the relationship between the quality of the results changes with the number of single photon sources used to generate the state. The figure of merit for quantifying the quality of these results is determined as the difference between the probability $p$ of generating the state and the quantum angle $QA\equiv \cos^{-1}\sqrt{\mathcal{F}}$~\cite{LI2015158}, which serves as the measure of the distance between the generated and target state. 
Our analysis reveals that the quality of the results consistently improves as the number of single photon sources employed increases. Indeed, these two quantities are positively correlated, with a Pearson correlation coefficient of approximately $92\%$ the sample Pearson correlation coefficient between the variable $\bm{x}$ and $\bm{y}$, $r_{x,y}\simeq92\%$, where the coefficient $r_{x,y}$ is defined as

\begin{equation}
    r_{x,y}=\frac{\sum_i(x_i-\bar{x})(y_i-\bar{y})}{\sqrt{\sum_i(x_i-\bar{x})^2}\sqrt{\sum_i(y_i-\bar{y})^2}},
\end{equation}
$\bar{x}$ is the sample mean of $\bm{x}$, while $x_i$ and $y_i$ are respectively the number of single photon sources, and the difference between probability and QA of the $i$-th optimization. 

We close with a brief discussion of single-photon sources on or compatible with integrated photonic devices. Currently, these have efficiencies above 84\% that are constantly improving via advances such as better mode coupling \cite{doi:10.1126/sciadv.abc8268}. They have high purities (99.3\%) and indistinguishabilities (98\%) \cite{doi:10.1126/sciadv.abc8268}. Together, purities and indistinguishabilities may each multiply our fidelities and the efficiencies may drop our success probabilities by a factor of $0.84^m$ for $m$ single-photon sources. Even so, the states generated using these noisy single-photon sources will be better than what is possible with ideal GBS devices alone, with $1-\mathcal{F}$ and $p$ on the last lines of Tables~\ref{tab:CATResults2}~\ref{tab:CATResults3}, and ~\ref{tab:GKPResults} being updated to $0.12< 0.32$ and $17\%>11\%$,  $8.7\times10^{-2}$ and $27\%>10\%$, and $1.1\times10^{-1}$ and $23\%>8.5\%$, respectively. Moreover, adding realistic noise only on the single-photon sources elicits a new tradeoff, with even cat states of amplitude $\alpha=2$ having higher overall success probabilities with one single-photon source than with two. As single-photon-source qualities improve including via quantum memories and as realistic noise sources such as loss on ideal GBS devices are considered, we expect the optimal results to converge to following the trends in Fig.~\ref{fig:ResultsvsSPSs3modes}.

\section{Conclusions}
We have introduced and evaluated an alternative scheme to non-Gaussian state generation with GBS-like devices presented in~\cite{PhysRevA.100.052301}. Specifically, we have compared the performance of the source when single photons and inline squeezing are employed in place of squeezed vacuum states in the input modes, serving as additional non-Gaussian resources for the PNRDs. Fidelity with the target state and probability of success of the conditional source have been used as figures of merit of the performance. We tested the efficiency of the model on 2- and 3-mode devices by targeting the Schr\"odinger cat state and the GKP code state, as two prominent examples of non-Gaussian states that complement Gaussian resources to realize universal quantum computation in the continuous-variable domain. 
Despite a variability in the results, ascribable to the dependence of the classical optimization on the initial conditions of the quantum circuit, overall, we observe that the introduction of single photons successfully increases the probability of generating the states as well as their fidelity with the target.
The 4-photon GKP core state $\ket{0_4}$, can be generated with fidelities greater than $90\%$ only when at least one squeezed photon interferes with the other mode. Moreover, the probability of generating a state with a fidelity above this threshold appears to be 8 times larger when two single photon sources are employed instead of one.   
Similarly, looking at the optimal probability of generating states whose fidelity with cat states of amplitude $\alpha=2$ is approximately equal to $95\%$, we see that it increases monotonically with the number of single photons introduced. 
Analogous evaluations with higher fidelities or more complex targets could be conducted with additional simulation resources, enabling the study of schemes with modes and measuring patterns that account for a greater number of photons. 
The advantage of the presented scheme over a traditional one can also be experimentally assessed using high efficiency single-photon sources and inline squeezing. Since state generation is used numerous times at the heart of quantum computation protocols, the advantages presented here should compound to make a marked difference in reducing the overhead for fault-tolerant quantum computation.

\begin{acknowledgments}
The authors acknowledge that the NRC headquarters is located on the traditional unceded territory of the Algonquin Anishinaabe and Mohawk people. The authors thank Guillaume Thekkadath and Anaelle Hertz for fruitful discussions and Filippo Miatto and Xanadu for technical support on their software packages. AZG acknowledges funding from the NSERC PDF program. KH acknowledges funding from NSERC's Discovery Grant.
\end{acknowledgments}

\clearpage

\onecolumngrid
\appendix
\section*{Supplementary Material}

\section{Single Photon Source}
\label{appendix:SPS}

\begin{figure}[htbp]
\begin{center}
    \includegraphics[trim=40 50 40 50, clip,width=0.5\textwidth]{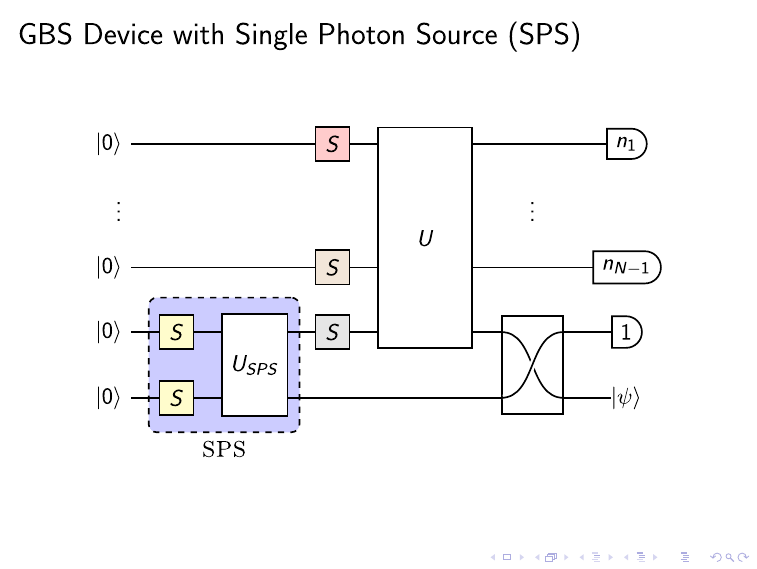}
\caption{GBS device with single photon source. The SPS is necessary because single photon Fock state are not handled by the optimizer.}
\label{fig:GBSwSPS}
\end{center}
\end{figure}

The Python libraries Strawberry Fields \cite{Killoran2019strawberryfields,Bromley_2020} and Mr Mustard~\cite{XanaduAI}, developed by Xanadu and used in this work, are indeed effective to simulate GBS devices. Non-Gaussian input states are however often mishandled by the logic of those libraries due to the inevitable need to truncate Fock spaces. As such, it is convenient using always squeezed vacuum states in the input modes of the interferometer. To overcome this problem, whenever a single photon Fock state $\ket{1}$ is chosen as input state, another 2-mode GBS is introduced as source of the Fock state $\ket{1}$. The two input states for this source are indeed Gaussian as well and the PNRD on the ancillary mode can be simulated when all the other modes of the main GBS are detected. In this way the ancillary GBS single photon source can be integrated with the actual GBS device under investigation. The simulator then effectively treats an $n$-mode GBS-device with $m$ single photon input states $\ket{1}$ as an $(n+m)$-mode linear interferometer where only one mode is left unmeasured, and the optimization is restricted to the tunable parameters of the main GBS device leaving the ancillary mode parameters fixed. An example with an $N$-mode GBS and a single photon source is given in Fig~\ref{fig:GBSwSPS}.

The Single Photon Source (SPS) shown in the violet box of Fig~\ref{fig:GBSwSPS} is created itself with another GBS device.
A Gaussian state can be prepared by squeezing single mode vacuum states and letting them interfere in a linear interferometer.
The single mode squeeze operator $\bm{S}(r)$ with $r\in\mathbb{R}$ is expressed in terms of creation operator $\hat{a}^\dagger$ and annihilation operator $\hat{a}$ as
\begin{equation}
    \bm{S}(r)=\exp\left[\frac{r}{2}\left(\hat{a}^2-\hat{a}^{\dagger2}\right)\right].
\end{equation}
The squeezed vacuum state $\ket{r}$ is then given by
\begin{equation}
    \ket{r}=\bm{S}(r)=\sum_{n=0}^\infty\alpha_n(r)\ket{2n},
\end{equation}
where
\begin{equation}
    \alpha_n(r)=\frac{\sqrt{(2n)!}}{2^nn!}\frac{\tanh^n{r}}{\sqrt{\cosh{r}}}.
\end{equation}
As it has been shown above any linear $N$-mode linear interferometer can be decomposed into $N(N-1)/2$ beam splitters $\bm{\theta,\varphi}$ and phase shift gate $\bm{R}(\phi)$.
The beam splitter operation in terms of creation  and annihilation operator on the modes $i$, and $j$ is
\begin{equation}\label{eq:BeamSplitter2}
    \bm{B}_{ij}(\theta,\varphi)=\exp\left[\theta\left(e^{i\varphi}\hat{a}_i\hat{a}_j^\dagger-e^{-i\varphi}\hat{a}_i^\dagger\hat{a}_j\right)\right]
\end{equation}

\begin{figure}[htbp]
\begin{center}
    \includegraphics[trim=10 580 10 40, clip,width=0.8\textwidth]{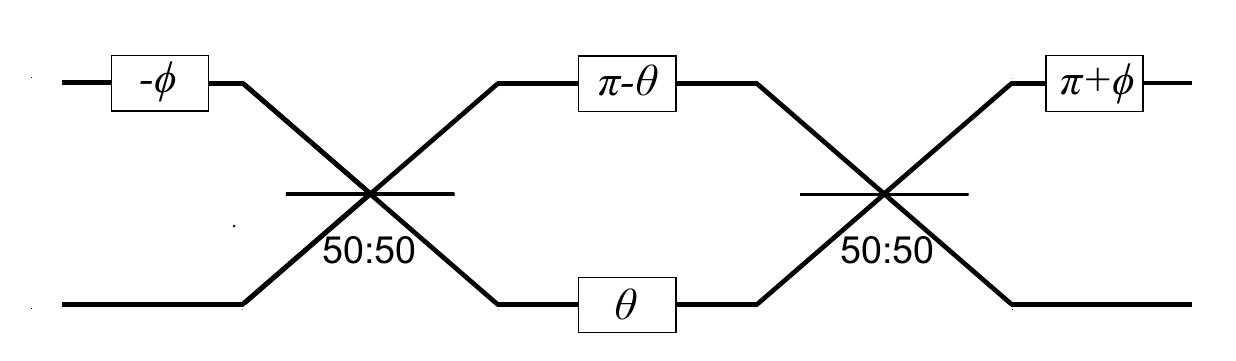}
\caption{Interferometer made of two symmetric 50:50 beam splitters and four phase shifters that realizes the evolution described in~\eqref{eq:UnitaryBS}}
\label{fig:BS_MZI}
\end{center}
\end{figure}

The unitary operator describing the evolution of the operators $\hat{a}_i,\hat{a}_j$  under the action of the beam splitter in $\eqref{eq:BeamSplitter2}$ is 
\begin{equation}\label{eq:UnitaryBS}
    \bm{U}(\theta,\varphi)=
    \begin{pmatrix}
        \cos{\theta} & e^{i\phi}\sin{\theta}\\
        -e^{-i\phi}\sin{\theta} & \cos{\theta}
    \end{pmatrix}
\end{equation}
where $\varphi=\pi-\phi$.
This evolution can be realized with an interferometer made of two $50:50$ beam splitters and 4 phase shifters arranged as displayed in Fig.~\ref{fig:BS_MZI}.

Given a two-mode pure input Fock state $\ket{n}\ket{m}$, the output state produced by the beam splitter $\bm{B}(\theta,\phi)$ is~\cite{agarwal_2012}
\begin{multline}
    \bm{B}(\theta,\phi)\ket{n}\ket{m}=\sum_{q,q'}\binom{n}{q}\binom{m}{q'}\sqrt{\frac{(q+q')!(n+m-q-q')!}{n!m!}}\\
    \times\cos(\theta)^{m+q-q'}\sin(\theta)^{n-q+q'}e^{i(n-q-q')\phi}(-1)^{q'}\ket{q+q',n+m-q-q'}
\end{multline}

We can now see how the two mode Gaussian state $\ket{\psi}$ produced by interfering two single mode squeezed vacuum states in the interferometer displayed in Fig.~\ref{fig:BS_MZI} can be expanded in the Fock basis. The 1 (2) label marks the first (second) mode
\begin{multline}
    \ket{\psi}=\bm{B}(\theta,\phi)\bm{S}_2(r_2)\bm{S}_1(r_1)\ket{0}_1\ket{0}_2=\sum_{n=0}^\infty\alpha_n(r_1)\bm{B}(\theta,\phi)\bm{S}_2(r_2)\ket{2n}_1\ket{0}_2=\sum_{n,m=0}^\infty\alpha_n(r_1)\alpha_m(r_2)\bm{B}(\theta,\phi)\ket{2n}\ket{2m}=\\
    =\sum_{n,m=0}^\infty\alpha_n(r_1)\alpha_m(r_2)\sum_{q,q'}\binom{n}{q}\binom{m}{q'}\sqrt{\frac{(q+q')!(2n+2m-q-q')!}{n!m!}}\\\times\cos(\theta)^{2m+q-q'}\sin(\theta)^{2n-q+q'}e^{i(2n-q-q')\phi}(-1)^{q'}\ket{q+q',2n+2m-q-q'}
\end{multline}
We know that a single photon state can be generated by measuring an odd number of photons in the output mode. In particular, a single photon Fock state whose stellar rank is equal to one can be produced even when only one photon is measured in the ancillary mode. In this case the unnormalized heralded state is


 \begin{multline}\label{eq:unnormalizedstate}
     \bra{1}_2\ket{\psi}=\bra{1}_2\bm{B}(\theta,\phi)\bm{S}_2(r_2)\bm{S}_1(r_1)\ket{0}_1\ket{0}_2\\
     =2\sum_{n,m=0}^\infty\frac{\tanh^n(r_1)\tanh^m(r_2)}{\sqrt{\cosh(r_1)\cosh(r_2)}}\frac{\sqrt{(2n+2m-1)!}}{2^{n+m}n!m!}(\cos(\theta))^{2n+1}(e^{-i\phi}\sin(\theta))^{2m-1}(n\tan^2(\theta)-m)\ket{2(n+m)-1}\\
     =\frac{2e^{i\phi}\cot{\theta}}{\sqrt{\cosh(r_1)\cosh(r_2)}}\sum_{N=1}^\infty\frac{\sqrt{(2N-1)!}}{2^N}\tanh^N(r_2)(e^{-i\phi}\sin{\theta})^{2N}\sum_{n=0}^N\left(\frac{\tanh{r_1}}{\tanh{r_2}}\right)^n\frac{(e^{i\phi}\cot{\theta})^{2n}}{n!(N-n!)}\left(\frac{n}{\cos^2(\theta)}-N\right)\ket{2N-1}
 \end{multline}
The single mode unnormalized state $\bra{1}_2\ket{\psi}$ can thus be written as
\begin{equation}
    \bra{1}_2\ket{\psi}=\mathcal{N}\sum_{N=1}^\infty c_N\ket{2N-1},
\end{equation}
where $\mathcal{N}=\frac{2e^{i\phi}\cot{\theta}}{\sqrt{\cosh(r_1)\cosh(r_2)}}$, and $c_N$ is the coefficient relative to the Fock state $\ket{2N-1}$.
In order to determine which values of $r_1,r_2,\theta,$ and $\phi$ lead to a single photon Fock state generation, we have to ensure that $c_1\neq0$, and $c_N=0~\forall N\geq2$

In general, we have that 
\begin{equation}\label{eq:cN}
    c_N \propto \sum_{n=0}^N\left(\frac{\tanh{r_1}}{\tanh{r_2}}\right)^n\frac{(e^{i\phi}\cot{\theta})^{2n}}{n!(N-n!)}\left(\frac{n}{\cos^2(\theta)}-N\right).
\end{equation}
By introducing the function $f(r_1,r_2)\equiv\left(\frac{\tanh{r_1}}{\tanh{r_2}}\right)$ we find that 
\begin{equation}\label{eq:c1}
    c_1 \propto  -1 + f(r_1,r_2)e^{2i\phi}
\end{equation}
while 
\begin{equation}\label{eq:c2}
    c_2 \propto  -1 + f(r_1,r_2)e^{2i\phi}(1-\cot^2(\theta))+f^2(r_1,r_2)e^{4i\phi}\cot^4(\theta).
\end{equation}

We proceed by setting $c_2=0$. We can prove by contradiction that $\phi\in\{\pi/2,3\pi/2\}$ is a necessary condition.
\begin{proof}
Let us assume that there exists $\phi\notin\{\pi/2,3\pi/2\}$, that satisfies $c_2=0$. If $c_2=0$, then both $\Re{c_2}=0$, and  $\Im{c_2}=0$
\begin{align}
    \Im{c_2}=0&\Rightarrow\sin(2\phi)f(r_1,r_2)(1-\cot^2(\theta)+2f(r_1,r_2)\cot^2(\theta)\cos(2\phi))=0\label{eq:im=0}\\ &\Rightarrow 1-\cot^2(\theta)+2f(r_1,r_2)\cot^2(\theta)\cos(2\phi)=0\\
    &\Rightarrow f(r_1,r_2) = \frac{\cot^2(\theta)-1}{2\cot^2(\theta)\cos(2\phi)} ;\\
    \Re{c_2}=0&\Rightarrow \frac{\cot^2(\theta)-1}{2\cot^2(\theta)\cos(2\phi)}\cos(2\phi)(1-\cot^2(\theta)) + \frac{(\cot^2(\theta)-1)^2}{4\cot^4(\theta)\cos^2(2\phi)}\cot^2(\theta)(2\cos^2(2\phi)-1)=1\\
    &\Rightarrow -\frac{(1-\cot^2(\theta))^2}{2\cot^2(\theta))}+\frac{(1-\cot^2(\theta))^2}{2\cot^2(\theta))}-\frac{(1-\cot^2(\theta))^2}{4\cot^4(\theta))\cos^2(2\phi)}=1\\
    &\Rightarrow 1= -\frac{(1-\cot^2(\theta))^2}{4\cot^4(\theta))\cos^2(2\phi)}\leq 0.
\end{align}
This proves that there does not exist any $\phi\notin\{\pi/2,3\pi/2\}$ that satisfies the condition $c_2=0$. While \eqref{eq:im=0} is verified only when $\sin(2\phi)=0$, and $\phi\in\{\pi/2,3\pi/2\}$.
\end{proof}

From \eqref{eq:cN} we can observe that the sign $e^{in\pi}=(-1)^n$ can be compensated by a change of sign of $r_1$ since $\tanh(r_1)$ is an odd function. As a consequence, we can set $\phi=\pi/2$ without loss of generality.

In this case we find that
\begin{align}
    c_2=0&\Rightarrow f(r_1,r_2)(1-\cot^2(\theta)+2f(r_1,r_2)\cot^2(\theta))=0\\
    &\Rightarrow 1  + f(r_1,r_2) = f(r_1,r_2)\cot^2(\theta)(1  + f(r_1,r_2))\label{eq:f_r1r2} .
\end{align}
Equation~\eqref{eq:f_r1r2} is verified when $f(r_1,r_2))=-1$. However, this condition also set $c_1$ in \eqref{eq:c1} to zero. The alternative solution is
\begin{equation}\label{eq:r1r2totheta}
    f(r_1,r_2)=\tan^2(\theta).
\end{equation}

This condition satisfies $c_N=0$ for any $N\geq2$. 
\begin{proof}
\begin{multline}
    c_N\propto\sum_{n=0}^N\frac{(-1)^n}{n!(N-n!)}\left(\frac{n}{\cos^2(\theta)}-N\right)=\sum_{n=0}^N\frac{(-1)^n}{n!(N-n!)}\frac{n}{\cos^2(\theta)}-N\sum_{n=0}^N\frac{(-1)^n}{n!(N-n!)}\\
    =\sum_{n=0}^{M+1}\frac{(-1)^n}{n!(M+1-n!)}\frac{n}{\cos^2(\theta)} -N\sum_{n=0}^N\frac{(1)^{N-n}(-1)^n}{n!(N-n!)}=\sum_{n=1}^{M+1}\frac{(-1)(-1)^{n-1}}{n(n-1)!(M-(n-1)!)}\frac{n}{\cos^2(\theta)}-N(1-1)^N=\\
    =-\frac{1}{\cos^2(\theta)}\sum_{m=0}^{M}\frac{(-1)^m}{m!(M-m!)}=-\frac{1}{\cos^2(\theta)}\sum_{m=0}^{M}\frac{(1)^{M-m}(-1)^m}{m!(M-m!)}=-\frac{1}{\cos^2(\theta)}(1-1)^M = 0 \quad \forall M\geq1.
\end{multline}
\end{proof}
The squeezing intensities $r_1,$ and $r_2$ are related with the beam splitter angle $\theta$ via the relation in \eqref{eq:r1r2totheta}. They can be optimized to maximize the probability $P(r_1,r_2)$ of generating the single photon Fock state. From \eqref{eq:unnormalizedstate} we have
\begin{align}
   P(r_1,r_2)&= \abs{\bra{1}_1\bra{1}_2\bm{B}\left(\theta,\frac{\pi}{2}\right)\bm{S}_2(r_2)\bm{S}_1(r_1)\ket{0}_1\ket{0}_2}^2=\abs{\frac{i\sin(2\theta)}{2\sqrt{\cosh(r_1)\cosh(r_2)}}(\tanh(r_1)+\tanh(r_2))}^2\\
   &=\abs{\frac{2\tan(\theta)}{1+\tan^2(\theta)}\frac{\tanh(r_1)+\tanh(r_2)}{2\sqrt{\cosh(r_1)\cosh(r_2)}}}^2=\abs{\frac{\sqrt{\frac{\tanh{r_1}}{\tanh{r_2}}}}{1+\frac{\tanh{r_1}}{\tanh{r_2}}}\frac{\tanh(r_1)+\tanh(r_2)}{\sqrt{\cosh(r_1)\cosh(r_2)}}}^2\\
    &=\abs{\frac{\sqrt{\sinh(r_1)\sinh(r_2)}}{\cosh(r_1)\cosh(r_2)}}^2=\abs{\frac{\sinh(r_1)\sinh(r_2)}{\cosh^2(r_1)\cosh^2(r_2)}}.
\end{align}
So $P(r_1,r_2)$ is a multiplicatively separable function with respect to the variables $r_1$, and $r_2$. Because of the symmetry of the function, a maximum is found for $r_1=r_2=r^*, \theta=\pi/4$, and $r^*=\text{arctanh}(1/\sqrt{2})\approx0.8814$ that maximizes the function $\frac{\sinh(r)}{\cosh^2(r)}$. In this case the probability of generating a single photon Fock state is $P(r^*,r^*)=1/4$.

\end{document}